\documentclass[pdflatex,sn-mathphys-num]{sn-jnl}

\usepackage{graphicx}%
\usepackage{multirow}%
\usepackage{amsmath,amssymb,amsfonts}%
\usepackage{amsthm}%
\usepackage[mathscr]{eucal}
\usepackage[title]{appendix}%
\usepackage{xcolor}%
\usepackage{textcomp}%
\usepackage{manyfoot}%
\usepackage{booktabs}%
\usepackage{algorithm}%
\usepackage{algorithmicx}%
\usepackage{algpseudocode}%
\usepackage{listings}%

\usepackage{amsthm}
\usepackage{lmodern} 
\usepackage{amsmath} 
\usepackage{fix-cm}
\fontsize{6}{7}\selectfont

\UseRawInputEncoding

\usepackage{afterpage}

\begin{document}

\title{Virtual Reality in Teacher Education: Insights from Pre-Service Teachers in Resource-limited Regions}


\author{\fnm{Matthew} \sur{Nyaaba}}\email{matthew.nyaaba@uga.edu}
\author{\fnm{Bisamrk Nyaaba} \sur{Akanzire}}\email{akanzire@gmail.com}
\author{\fnm{Macharious} 
\sur{Nabang}}\email{machariousnabang@gmail.com}
\affil{\orgdiv{Department of Educational Theory and Practice}, \orgname{University of Georgina}, \city{Athens}, \state{Georgia}, \country{USA}}

\affil{\orgdiv{Department of Education}, \orgname{Gambaga College of Education}, \city{Gambaga}, \state{N/E}, \country{Ghana}}

\affil{\orgdiv{Creative Arts Department}, \orgname{Bagabaga College of Education}, \city{Tamale}, \state{N/R}, \country{Ghana}}

\abstract{This study explores the perceptions, challenges, and opportunities associated with using Virtual Reality (VR) as a tool in teacher education among pre-service teachers in a resource-limited setting. The research specifically addresses (1) how participation in a VR-facilitated lesson influences pre-service teachers' perceptions of VR in education and (2) the unique challenges and opportunities identified by these pre-service teachers regarding VR's use in education. Utilizing a qualitative case study design, the study draws on the experiences and reflections of 36 Ghanaian pre-service early childhood educators who engaged with VR in a facilitated lesson for the first time.
Findings reveal that initial exposure to VR generated a positive perception, with participants highlighting VR's potential as an engaging and interactive tool that can support experiential learning. Notably, many participants saw the VR-facilitated lesson as a promising alternative to synchronous online learning, particularly for its ability to simulate in-person presentations. They believe VR's immersive capabilities could enhance both teacher preparation and learner engagement in ways that traditional teaching often does not, especially noting that VR has the potential of addressing expensive educational field trips. Despite these promising perceptions, participants identified key challenges, including limited infrastructure, unreliable internet connectivity, and insufficient access to VR equipment as perceived challenges that might hinder the integration of VR in a resource-limited region like Ghana.
These findings offer significant implications for educational policymakers and institutions aiming to leverage VR to enhance teacher training and professional development in similar contexts to consider addressing the perceived challenges for successful VR integration in education. We recommend further empirical research be conducted involving pre-service teachers use of VR in their classrooms.}




\maketitle

\section{Introduction}\label{sec1}

As a rapidly advancing educational tool, virtual reality (VR) holds significant potential to transform teaching practices, especially within teacher education. With its immersive and interactive capabilities, VR fosters experiential learning that extends beyond traditional instructional methods, allowing pre-service teachers to engage deeply with realistic, simulated teaching environments \cite{Burns2012, Lopez2023}. In teacher education, this interactivity can be particularly impactful, as VR enables pre-service teachers to navigate complex classroom dynamics and apply pedagogical strategies in a risk-free, controlled setting, enhancing their engagement and preparedness for real-world teaching \cite{Fernandez2019, Perez2021}. However, despite the promise VR holds, its application in teacher education within developing countries, such as Ghana, remains underdeveloped, with limited exposure and resources to support its implementation \cite{Oyelere2020}.

The limited application of VR in Ghanaian teacher training reveals critical gaps, including a lack of foundational exposure to immersive technology, insufficient infrastructure, and an absence of systematic training programs to support educators in integrating VR meaningfully into their practices \cite{Boafo2024, Narh2019}. While existing studies underscore VR's benefits in enhancing engagement and competency development among trainee teachers \cite{Asare2023}, they also highlight that these benefits are contingent on institutional support, cultural relevance, and the availability of technical resources \cite{Morales2023}. These barriers are particularly pronounced in resource-limited settings, where access to reliable internet and VR equipment is often constrained, posing challenges to VR's sustained use and effectiveness \cite{Asare2023, Boafo2024}.

This study seeks to explore the perceptions and perceived challenges of pre-service teachers in resource-limited regions who participated in a VR-facilitated lesson for the first time. In doing so, the study contributes to the growing body of research on VR in education, highlighting critical considerations for expanding VR's accessibility and effectiveness in global teacher training contexts. The study was guided by the following research questions:

\begin{enumerate}
    \item How does participation in a VR-facilitated lesson influence pre-service teachers' perceptions of VR in education in resource-limited regions?
    
    \item What key challenges and opportunities do pre-service teachers perceive in using VR in education in resource-limited regions?
\end{enumerate}

\section{Recent Studies on VR in Education }
Recent research on VR in education has illuminated both the possibilities and the challenges of employing immersive technology for professional development in teaching. For instance, a study by \cite{Fernandez2019} confirms this potential, showing how immersive environments support skill development and reflective practice among teachers, particularly in classroom management and engagement techniques. Similar findings are presented by \cite{Perez2021}, who explored VR and augmented reality (AR) as tools for university-level teacher training, noting these technologies' ability to foster experiential learning that translates to more effective instructional strategies. Likewise, current research from Ghanaian colleges of education by \cite{Asare2023} highlights that interactive technology, although limited in use, offers significant benefits in enhancing engagement and learning outcomes in physical education (PE).

While tools like interactive whiteboards and fitness trackers show promise for tracking progress and improving classroom interaction, their integration remains low due to resource constraints, lack of teacher training, and institutional barriers \cite{Asare2023}. These studies reveal a core advantage of VR in teacher training: it enables experiential, simulated learning that fosters professional growth by allowing teachers to interact in realistic educational scenarios, thus bridging theoretical knowledge with practice. Importantly, researchers emphasize the need for strategic implementation, and \cite{Rojas2020} recommends that institutions adopting VR should provide the necessary support structures and training for educators to maximize VR's efficacy.

\section {Study Approach}
This study utilized a case study design to explore how VR technology can impact pre-service teachers' perceptions and perceived challenges in resource-limited context. This design allowed for a close examination of this unique group's experiences and providing in-depth insights into VR's potential as a teaching tool in their context.
\subsection{Participants }
Participants included 36 Early Childhood pre-service teachers and Visual Arts pre-service teachers from two colleges of education in Ghana. The colleges selected involved a premium and a recently established college of education being the home colleges of the researchers. The study involved students who participated in VR-Facilitated Guest Lecture on Technology in Education by a presenter from the University of Georgia and were willing to share their views and reflections for this study. The dataset highlights diverse participant demographics, such as age ranges (e.g., 20-25, 26-30). 
\subsection{Data Collection }\label{sec3}
Data collection extended over a series of interactive and reflective sessions. Following the VR presentation, focus group discussions facilitated by the instructor created a platform for participants to share reflections, ask questions, and explore VR's applications in teaching. These discussions allowed participants to voice initial impressions, common challenges (e.g., technical difficulties), and perceived educational benefits through the focus group and reflective journal using Google Form. The focus group discussions provided valuable qualitative data on shared challenges and collective insights into VR's role in enhancing early-grade education, as well as its potential limitations within the resources-limited region, like Ghanaian. In the data collection process, participants were without VR headsets but joined the session via the Portal Screen, allowing them to view the VR environment through a screen projection. This setup enabled real-time interaction between the VR presenter and non-VR participants, facilitating an inclusive learning experience.  (See Figure 1 and 2). The topic of the presentation was about VR in education, potential and challenges. 

\begin{figure}[H]
\centering
\includegraphics[width=\textwidth]{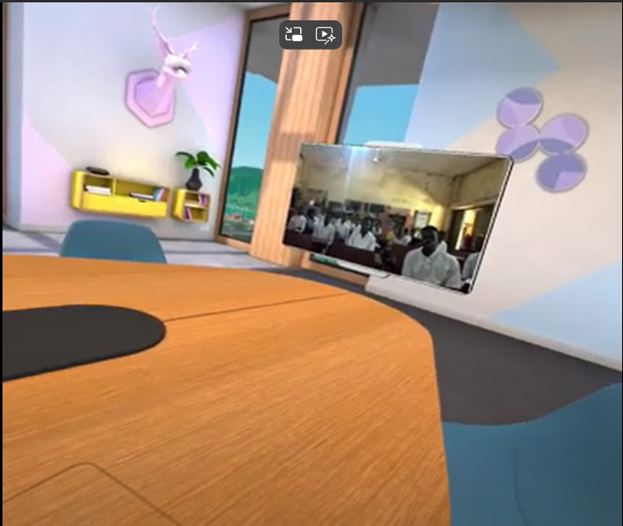} 
\caption{Participants in the Horizon Workroom Portal Screen}
\label{fig:Picture1.png}
\end{figure}

\begin{figure}[H]
\centering
\includegraphics[width=\textwidth]{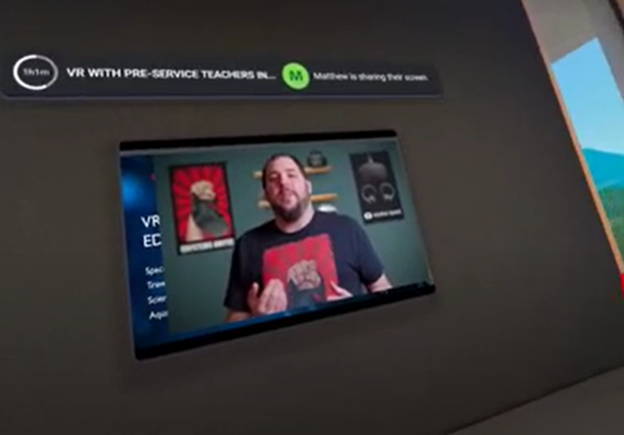} 
\caption{: Video showing in the VR Horizon Workroom }
\label{fig:Picture2.png}
\end{figure}

\section{Results and Discussion}
The data for this study were analyzed using thematic analysis, focusing on insights gathered from focus group discussions and reflective journals. This approach allowed for an in-depth exploration of participants' perceptions, experiences, and challenges regarding VR's role in teacher education within a resource-limited regions. Common themes emerged, include VR's potential as an alternative to traditional synchronous online learning and its effectiveness in simulating realistic teaching scenarios.

\subsection{Participants' Knowledge and Initial Exposure to VR}\label{sec:knowledge-exposure}
A large portion of participants (27 out of 36) indicated this was their first experience with VR, underscoring its novelty in Ghanaian teacher training (See Figure 1). This lack of prior exposure highlights VR's potential to introduce teachers to immersive educational tools but also points to the need for foundational training to maximize its benefits. As one participant expressed, "The use of VR is very interesting and helpful," capturing the initial intrigue VR sparked among first-time users (Participant 1). Another noted, "VR is very good and helpful," further emphasizing the positive initial response to this new technology in an educational context (Participant 2).

For those with some prior exposure to VR, interactions were periodic and often encountered informally, such as "in school" or through specific events, rather than through structured training. This sporadic exposure reflects a lack of systematic VR integration within Ghanaian teacher education programs, potentially limiting VR's effectiveness. One participant remarked, "VR would be very much helpful to all students, so I think it should be made available and accessible to all educational institutions," voicing a desire for broader access and formal integration of VR in education (Participant 4). Another participant echoed this sentiment, stating, "It would be good for the government of each country to provide VR to all the schools in their country," suggesting the perceived importance of VR accessibility in formal education (Participant 6).

These findings align with \cite{Marougkas2023}, who emphasize that educational VR's effectiveness depends on prior exposure and familiarity with the technology. The significant portion of first-time VR users among the Ghanaian teacher trainees underscores the novelty of VR as a learning tool, supporting \cite{Huang2021} and \cite{Burns2012}, who argue that immersive VR can serve as a powerful pedagogical tool for teacher development, provided foundational training is in place. Participant feedback, expressing both excitement and a desire for more systematic VR integration, confirms the value of early, structured VR experiences in shaping positive user attitudes. However, sporadic prior exposure suggests that current integration within Ghana's teacher training is limited, a gap that \cite{Alvarez2024} and \cite{Asare2023} also identify as an obstacle to VR adoption in educational systems, particularly in emerging contexts.

\begin{figure}[H]
\centering
\includegraphics[width=\textwidth]{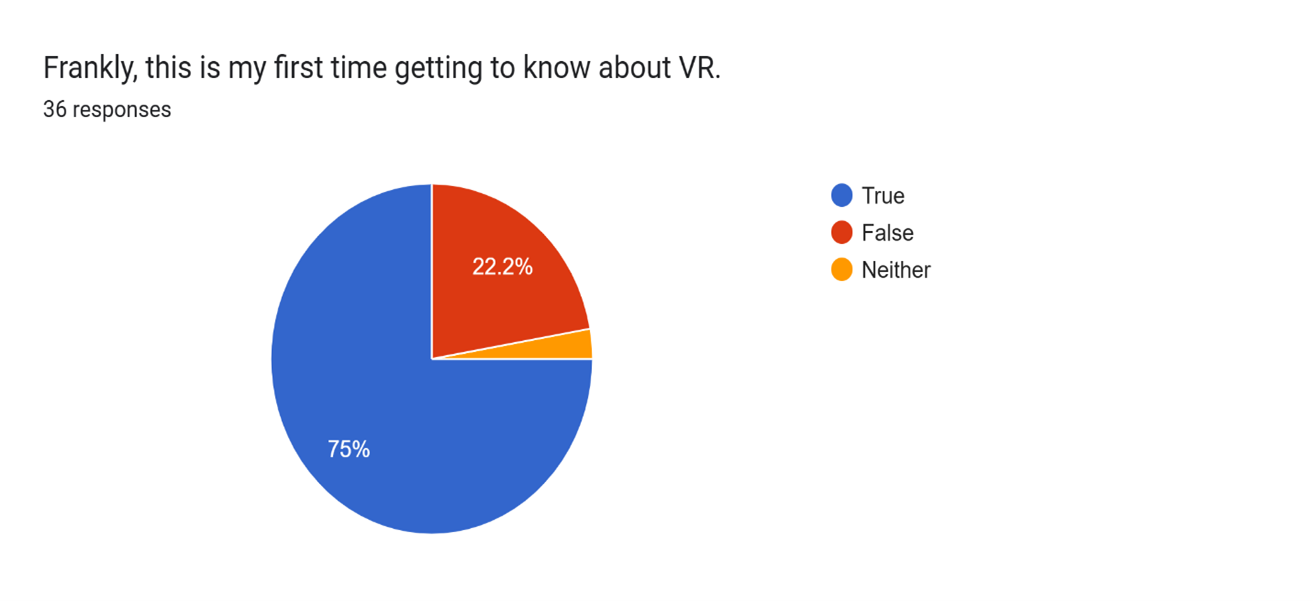} 
\caption{: Participants Awareness of VR}
\label{fig:Picture3.png}
\end{figure}

\subsection{Knowledge Enhancement and Enjoyment}
Participants widely perceived VR as an effective tool for enhancing content knowledge, with many seeing it as beneficial to their professional development. This aligns with one participant's observation, "VR will help learners to learn best in technology," highlighting the belief that VR can make complex concepts more understandable and engaging, especially with subjects that are mostly taught abstractly or require expensive field trips like visiting other parts of the world (Participant 5). Another participant specifically noted VR's value in early childhood education, saying, "VR is very good because it can help early graders to acquire knowledge, but what makes it difficult is the network" (Participant 8), underscoring both the potential benefits of VR and the technical challenges that may hinder its full integration.

The general enthusiasm by the participants of this study for VR's potential to enrich learning and make it more enjoyable supports findings by \cite{Oyelere2020} and \cite{Asare2023}, who identified VR's interactive elements as highly engaging for learners and contradicts \cite{aboyinga2020factors} studies that pre-service teachers lack motivation when virtual learning. The participants' positive experiences mirror \cite{Alvarez2024}, who found VR's interactive nature vital in cultivating an engaging learning environment. Furthermore, \cite{Marougkas2023} note that VR aligns well with constructivist and experiential learning theories, and participants' feedback, emphasizing the interactivity and accessibility of VR, supports this view.

\subsection{Engagement and Interaction Compared to Traditional Methods}\label{sec:engagement-interaction}
Participant responses on VR's interactivity compared to platforms like Zoom were mixed, indicating varying levels of comfort and familiarity with VR technology. Some participants valued VR's immersive qualities, with one remarking, "It helps the learner and the teacher to work together," highlighting VR's potential for collaborative learning beyond what traditional platforms might offer (Participant 11). However, others expressed concerns about VR's time demands, as one participant noted, "It seems we have been spending much time on this VR, yeah, and how am I going to benefit with the VR?" (Participant 12), reflecting reservations about VR's practical advantages relative to familiar tools.

These findings echo \cite{Chen2022}, who reported similar ambivalence, with some users finding VR highly engaging yet time-intensive. This reinforces \cite{Alvarez2024}, who suggest that user acceptance of VR in education hinges on a balance of immersive benefits with practical efficiency. Despite mixed responses, many participants showed sustained curiosity about VR, with one stating, "Visual reality can help boost teaching lessons and make it seem more real," signaling interest in VR's potential to enrich traditional educational methods (Participant 13). This aligns with \cite{Burns2012} and \cite{Marougkas2023}, who emphasize that VR's unique features can inspire educators to explore innovative teaching methods.

\subsection{Perceived Challenges of VR Implementation}\label{sec:challenges-implementation}
Technical challenges, particularly with internet connectivity, emerged as significant barriers to VR implementation, with participants frequently citing connectivity issues as a hindrance. One participant expressed, "VR is very good, but what makes it difficult is the network," reflecting the infrastructure limitations often encountered in developing regions (Participant 8). Another participant echoed this sentiment, noting, "VR is more helpful, but the challenge is poor internet connectivity," emphasizing the need for reliable technical support to sustain VR's use as an educational tool (Participant 15).

These insights reinforce findings by \cite{Alvarez2024} and \cite{Oyelere2020}, who identified technical infrastructure as a crucial factor in VR's successful implementation in education, especially in areas with limited resources. \cite{Marougkas2023} also note that the potential of VR is closely tied to technological robustness, as insufficient infrastructure can undermine VR's educational benefits. \cite{Burns2012} similarly cautions that while VR offers significant pedagogical advantages, practical challenges must be addressed to realize its full impact. The participants' feedback suggests that addressing infrastructural gaps will be essential for VR and the integration of many technological tools to achieve sustainability in teacher education in most resource-limited regions \cite{nyaaba2021challenges, nyaaba2021pre}.

\subsection{Conclusions}
This study highlights VR as a promising tool for enhancing teacher education, particularly in resource-limited contexts like Ghana. VR offers immersive and interactive experiences that allow pre-service teachers to engage with realistic teaching scenarios, which increases their preparedness for classroom challenges and improves their engagement with complex topics. Participants noted that VR could serve as a valuable alternative to traditional synchronous online learning, particularly because of its capacity to simulate in-person teaching interactions and support experiential learning. Additionally, VR can facilitate the teaching of topics, such as those in science, that often require field trips or hands-on exploration. However, the study also identifies significant barriers to VR's successful integration, including limited access to VR technology, unreliable internet connectivity, and a lack of foundational training, all of which restrict VR's effectiveness and sustainability in resource-limited educational settings.

\subsection{Recommendations}
To realize VR's potential in teacher education, institutions and policymakers should prioritize infrastructure investments to ensure reliable internet, VR equipment, and technical support, enabling consistent use of VR as an educational tool. Additionally, foundational training programs are essential to help pre-service teachers build technical proficiency and develop instructional strategies specific to VR's applications in teacher education in resource-limited regions. Given VR's immersive qualities, institutions might also consider using it as a complement or substitute for conventional online learning platforms like Zoom, particularly for interactive and hands-on practice. Developing locally relevant VR content that reflects Ghanaian cultural and educational contexts could further enhance its effectiveness. Finally, policymakers are encouraged to support VR's integration in teacher training through funding initiatives and partnerships with educational technology providers, helping to reduce costs and increase access. By addressing these needs, VR can be a transformative tool for teacher preparation, fostering more engaged, prepared, and adaptable educators.


\bibliography{sn-bibliography}


\begin{thebibliography}{17}
\ifx \bisbn   \undefined \def \bisbn  #1{ISBN #1}\fi
\ifx \binits  \undefined \def \binits#1{#1}\fi
\ifx \bauthor  \undefined \def \bauthor#1{#1}\fi
\ifx \batitle  \undefined \def \batitle#1{#1}\fi
\ifx \bjtitle  \undefined \def \bjtitle#1{#1}\fi
\ifx \bvolume  \undefined \def \bvolume#1{\textbf{#1}}\fi
\ifx \byear  \undefined \def \byear#1{#1}\fi
\ifx \bissue  \undefined \def \bissue#1{#1}\fi
\ifx \bfpage  \undefined \def \bfpage#1{#1}\fi
\ifx \blpage  \undefined \def \blpage #1{#1}\fi
\ifx \burl  \undefined \def \burl#1{\textsf{#1}}\fi
\ifx \doiurl  \undefined \def \doiurl#1{\url{https://doi.org/#1}}\fi
\ifx \betal  \undefined \def \betal{\textit{et al.}}\fi
\ifx \binstitute  \undefined \def \binstitute#1{#1}\fi
\ifx \binstitutionaled  \undefined \def \binstitutionaled#1{#1}\fi
\ifx \bctitle  \undefined \def \bctitle#1{#1}\fi
\ifx \beditor  \undefined \def \beditor#1{#1}\fi
\ifx \bpublisher  \undefined \def \bpublisher#1{#1}\fi
\ifx \bbtitle  \undefined \def \bbtitle#1{#1}\fi
\ifx \bedition  \undefined \def \bedition#1{#1}\fi
\ifx \bseriesno  \undefined \def \bseriesno#1{#1}\fi
\ifx \blocation  \undefined \def \blocation#1{#1}\fi
\ifx \bsertitle  \undefined \def \bsertitle#1{#1}\fi
\ifx \bsnm \undefined \def \bsnm#1{#1}\fi
\ifx \bsuffix \undefined \def \bsuffix#1{#1}\fi
\ifx \bparticle \undefined \def \bparticle#1{#1}\fi
\ifx \barticle \undefined \def \barticle#1{#1}\fi
\bibcommenthead
\ifx \bconfdate \undefined \def \bconfdate #1{#1}\fi
\ifx \botherref \undefined \def \botherref #1{#1}\fi
\ifx \url \undefined \def \url#1{\textsf{#1}}\fi
\ifx \bchapter \undefined \def \bchapter#1{#1}\fi
\ifx \bbook \undefined \def \bbook#1{#1}\fi
\ifx \bcomment \undefined \def \bcomment#1{#1}\fi
\ifx \oauthor \undefined \def \oauthor#1{#1}\fi
\ifx \citeauthoryear \undefined \def \citeauthoryear#1{#1}\fi
\ifx \endbibitem  \undefined \def \endbibitem {}\fi
\ifx \bconflocation  \undefined \def \bconflocation#1{#1}\fi
\ifx \arxivurl  \undefined \def \arxivurl#1{\textsf{#1}}\fi
\csname PreBibitemsHook\endcsname

\bibitem[\protect\citeauthoryear{Burns}{2012}]{Burns2012}
\begin{barticle}
\bauthor{\bsnm{Burns}, \binits{M.}}:
\batitle{Immersive learning for teacher professional development}.
\bjtitle{eLearn}
\bvolume{2012}(\bissue{4}),
\bfpage{1}
(\byear{2012})
\doiurl{10.1145/2181207.2181208}
\end{barticle}
\endbibitem

\bibitem[\protect\citeauthoryear{López}{2023}]{Lopez2023}
\begin{barticle}
\bauthor{\bsnm{López}, \binits{D.}}:
\batitle{Evaluation of the effectiveness of virtual reality in teacher training}.
\bjtitle{Educational Technology Research and Development}
\bvolume{68}(\bissue{5}),
\bfpage{2503}--\blpage{2520}
(\byear{2023})
\doiurl{10.1007/s11423-020-09742-8}
\end{barticle}
\endbibitem

\bibitem[\protect\citeauthoryear{Fernández}{2019}]{Fernandez2019}
\begin{barticle}
\bauthor{\bsnm{Fernández}, \binits{M.Ã.}}:
\batitle{Immersive environments and their effect on teacher professional development}.
\bjtitle{Computers \& Education}
\bvolume{139},
\bfpage{103755}
(\byear{2019})
\doiurl{10.1016/j.compedu.2019.103755}
\end{barticle}
\endbibitem

\bibitem[\protect\citeauthoryear{Pérez}{2021}]{Perez2021}
\begin{barticle}
\bauthor{\bsnm{Pérez}, \binits{A.M.}}:
\batitle{Immersive technologies for university teacher training}.
\bjtitle{Sustainability}
\bvolume{13}(\bissue{19}),
\bfpage{10740}
(\byear{2021})
\doiurl{10.3390/su131910740}
\end{barticle}
\endbibitem

\bibitem[\protect\citeauthoryear{Oyelere et~al.}{2020}]{Oyelere2020}
\begin{barticle}
\bauthor{\bsnm{Oyelere}, \binits{S.S.}},
\bauthor{\bsnm{Bouali}, \binits{N.}},
\bauthor{\bsnm{Kaliisa}, \binits{R.}}, \betal:
\batitle{Exploring the trends of educational virtual reality games: A systematic review of empirical studies}.
\bjtitle{Smart Learning Environments}
\bvolume{7},
\bfpage{31}
(\byear{2020})
\doiurl{10.1186/s40561-020-00142-7}
\end{barticle}
\endbibitem

\bibitem[\protect\citeauthoryear{Boafo et~al.}{2024}]{Boafo2024}
\begin{barticle}
\bauthor{\bsnm{Boafo}, \binits{F.}},
\bauthor{\bsnm{Dotse}, \binits{H.A.A.}},
\bauthor{\bsnm{Mensah}, \binits{S.}},
\bauthor{\bsnm{Boadu}, \binits{D.}}:
\batitle{Exploring the impact of virtual reality on stem education in ghana}.
\bjtitle{International Journal of Research and Scientific Innovation}
(\byear{2024})
\doiurl{10.51244/IJRSI.2024.1104034}
\end{barticle}
\endbibitem

\bibitem[\protect\citeauthoryear{Narh et~al.}{2019}]{Narh2019}
\begin{botherref}
\oauthor{\bsnm{Narh}, \binits{N.}},
\oauthor{\bsnm{Boateng}, \binits{R.}},
\oauthor{\bsnm{Afful-Dadzie}, \binits{E.}},
\oauthor{\bsnm{Owusu}, \binits{A.}}:
Virtual platforms: Assessing the challenges of e-learning in ghana.
Journal of Online Learning Research
(2019)
\end{botherref}
\endbibitem

\bibitem[\protect\citeauthoryear{Asare et~al.}{2023}]{Asare2023}
\begin{barticle}
\bauthor{\bsnm{Asare}, \binits{S.}},
\bauthor{\bsnm{Kyenkyehene}, \binits{S.A.}},
\bauthor{\bsnm{Emmanuel}, \binits{M.K.}}:
\batitle{Interactive technology in physical education classroom: A case of a ghanaian college of education}.
\bjtitle{American Journal of Education and Information Technology}
\bvolume{7}(\bissue{2}),
\bfpage{51}--\blpage{58}
(\byear{2023})
\doiurl{10.11648/j.ajeit.20230702.11}
\end{barticle}
\endbibitem

\bibitem[\protect\citeauthoryear{Morales}{2023}]{Morales2023}
\begin{barticle}
\bauthor{\bsnm{Morales}, \binits{T.}}:
\batitle{Virtual reality and its impact on higher education: An empirical study}.
\bjtitle{Education Sciences}
\bvolume{11}(\bissue{8}),
\bfpage{384}
(\byear{2023})
\doiurl{10.3390/educsci11080384}
\end{barticle}
\endbibitem

\bibitem[\protect\citeauthoryear{Rojas}{2020}]{Rojas2020}
\begin{barticle}
\bauthor{\bsnm{Rojas}, \binits{C.}}:
\batitle{Application of virtual reality in higher education: A case study}.
\bjtitle{Educational Technology Research and Development}
\bvolume{68}(\bissue{5}),
\bfpage{2503}--\blpage{2520}
(\byear{2020})
\doiurl{10.1007/s11423-020-09742-8}
\end{barticle}
\endbibitem

\bibitem[\protect\citeauthoryear{Marougkas et~al.}{2023}]{Marougkas2023}
\begin{barticle}
\bauthor{\bsnm{Marougkas}, \binits{A.}},
\bauthor{\bsnm{Troussas}, \binits{C.}},
\bauthor{\bsnm{Krouska}, \binits{A.}},
\bauthor{\bsnm{Sgouropoulou}, \binits{C.}}:
\batitle{Virtual reality in education: A review of learning theories, approaches and methodologies for the last decade}.
\bjtitle{Electronics}
\bvolume{12}(\bissue{13}),
\bfpage{2832}
(\byear{2023})
\doiurl{10.3390/electronics12132832}
\end{barticle}
\endbibitem

\bibitem[\protect\citeauthoryear{Huang et~al.}{2021}]{Huang2021}
\begin{barticle}
\bauthor{\bsnm{Huang}, \binits{Y.}},
\bauthor{\bsnm{Richter}, \binits{E.}},
\bauthor{\bsnm{Kleickmann}, \binits{T.}},
\bauthor{\bsnm{Richter}, \binits{D.}}:
\batitle{Virtual reality in teacher education from 2010 to 2020: A review of program implementation, intended outcomes, and effectiveness measures}.
\bjtitle{Education Sciences}
(\byear{2021})
\doiurl{10.35542/osf.io/ye6uw}
\end{barticle}
\endbibitem

\bibitem[\protect\citeauthoryear{Álvarez et~al.}{2024}]{Alvarez2024}
\begin{barticle}
\bauthor{\bsnm{Álvarez}, \binits{I.M.}},
\bauthor{\bsnm{Manero}, \binits{B.}},
\bauthor{\bsnm{Romero-Hernández}, \binits{A.}}:
\batitle{Virtual reality platform for teacher training on classroom climate management: Evaluating user acceptance}.
\bjtitle{Virtual Reality}
\bvolume{28},
\bfpage{78}
(\byear{2024})
\doiurl{10.1007/s10055-024-00973-6}
\end{barticle}
\endbibitem

\bibitem[\protect\citeauthoryear{Aboyinga and Nyaaba}{2020}]{aboyinga2020factors}
\begin{barticle}
\bauthor{\bsnm{Aboyinga}, \binits{J.}},
\bauthor{\bsnm{Nyaaba}, \binits{M.}}:
\batitle{Factors that ensure motivation in virtual learning among college of education students in ghana: The emergency remote teaching (ert) during covid'19 pandemic}.
\bjtitle{European Journal of Research and Reflection in Educational Sciences Vol}
\bvolume{8}(\bissue{9}),
\bfpage{1}--\blpage{9}
(\byear{2020})
\end{barticle}
\endbibitem

\bibitem[\protect\citeauthoryear{Chen}{2022}]{Chen2022}
\begin{barticle}
\bauthor{\bsnm{Chen}, \binits{C.-Y.}}:
\batitle{Immersive virtual reality to train preservice teachers in managing students' challenging behaviours: A pilot study}.
\bjtitle{British Journal of Educational Technology}
(\byear{2022})
\doiurl{10.1111/bjet.13181}
\end{barticle}
\endbibitem

\bibitem[\protect\citeauthoryear{Nyaaba and Sandawey}{2021}]{nyaaba2021challenges}
\begin{barticle}
\bauthor{\bsnm{Nyaaba}, \binits{M.}},
\bauthor{\bsnm{Sandawey}, \binits{S.}}:
\batitle{Challenges and attitude towards blended learning among college of education students in ghana}.
\bjtitle{Journal of Education and Practices ISSN 2617-5444 (ONLINE) \& ISSN 2617-6874 (PRINT)}
\bvolume{3}(\bissue{2}),
\bfpage{41}--\blpage{47}
(\byear{2021})
\end{barticle}
\endbibitem

\bibitem[\protect\citeauthoryear{Nyaaba et~al.}{2021}]{nyaaba2021pre}
\begin{barticle}
\bauthor{\bsnm{Nyaaba}, \binits{M.}},
\bauthor{\bsnm{Aboyinga}, \binits{J.}},
\bauthor{\bsnm{Akanzire}, \binits{B.N.}}:
\batitle{Pre-service parents teachers' attitude and perceived challenges about inclusive education in ghana: The ghanaian inclusive education policy}.
\bjtitle{American Journal of Educational Research}
\bvolume{9}(\bissue{6}),
\bfpage{341}--\blpage{346}
(\byear{2021})
\end{barticle}
\endbibitem

\end{thebibliography}

\end{document}